# Interaction force in a vertical dust chain inside a glass box

Jie Kong, Ke Qiao, Lorin S. Matthews and Truell W. Hyde

Center for Astrophysics, Space Physics, and Engineering Research (CASPER)

Baylor University

Waco, Texas 76798-7310

Abstract

Small number dust particle clusters can be used as probes for plasma diagnostics. The number of dust particles as well as cluster size and shape can be easily controlled employing a glass box placed within a GEC rf reference chamber to provide confinement of the dust. The plasma parameters inside this box and within the larger plasma chamber have not yet been adequately defined. Adjusting the rf power alters the plasma conditions causing structural changes of the cluster. This effect can be used to probe the relationship between the rf power and other plasma parameters. This experiment employs the sloshing and breathing modes of small cluster oscillations to examine the relationship between system rf power and the particle charge and plasma screening length inside the glass box. The experimental results provided indicate that both the screening length and dust charge decrease as rf power inside the box increases. The decrease in dust charge as power increases may indicate that ion trapping plays a significant role in the sheath.

1. Introduction

Complex plasmas, partially ionized gases containing small, usually micron-sized, dust particles, are of great interest in part because they provide an ideal and versatile analogue to the study of finite charged-particle systems. The dust particles collect ions and electrons from the plasma generally obtaining a negative charge due to the higher mobility of the electrons. In laboratory experiments on Earth, gravity causes the dust particles to settle within the plasma, where they are then levitated by the electrostatic field above the lower powered electrode which is also negatively charged within the plasma chamber. Utilizing a glass box placed on the lower electrode to provide strong horizontal confinement allows clusters with small numbers of particles to be easily manipulated, and multiple and/or single vertical chains to be formed. Previous experiments have shown that the structural transition between multiple vertical chains and a single vertical chain in the glass box is achieved by varying the system rf power. This transition is assumed to be due to the variation in the plasma ionization rate that occurs during such a change in system rf power [1]. The goal of this investigation is to experimentally explore the relationship between system rf power and global dusty plasma parameters such as the Debye length and dust charge.

The paper is structured as follows: Section 2 presents the theoretical background for the experiment and analysis. Section 3 describes the experimental setup and presents the results. In Section 4, a detailed analysis of the experimental data is provided. Finally, a discussion of these results is given in Section 5 while conclusions are given in Section 6.

2. Theoretical background

The total energy of an N-particle dust cluster within an anisotropic, horizontally directed confinement is given by [2 – 5],

$$U = \sum_{i=1}^{N} \frac{1}{2} m_d \omega_0^2 \left( \beta x_i^2 + y_i^2 \right) + \sum_{pairs} \frac{Q_d^2}{4\pi\varepsilon_0} \frac{e^{-\kappa r_{ij}}}{r_{ij}} \tag{1}$$

where $x_i$, $y_i$ are the horizontal and vertical coordinates respectively, $r_{ij}$ is the separation distance between the $i^{th}$ and $j^{th}$ particles, $m_d$ and $Q_d$ are the dust particle's mass and charge respectively, $\omega_0 = \omega_y$ is the frequency of oscillation in the vertical direction, $\beta = \omega_x^2/\omega_y^2$ is the anisotropy parameter ($\omega_x$ is the frequency of oscillation in the horizontal direction) and $\kappa = 1/\lambda_D$ is the shielding parameter. The Yukawa potential seen in the second term on the right hand side of Eq 1 represents the interaction between the dust particles. The equilibrium separation distance between dust particles $(R_{eq})$ and the relationship between their breathing and sloshing frequencies may be derived from the conditions ($U' = 0$ and $U''$ respectively).

In this experiment, a vertically aligned two-particle system was chosen to allow examination of the interparticle interaction. A two-particle system was chosen due to the fact that this structure provides the simplest form of Yukawa interaction and is therefore easiest to analyze. According to Eq 1, the total potential energy of a vertically aligned two-particle system $U_2$ is

$$U_2 = \frac{1}{2} m_d \omega_0^2 \left( \left(\frac{R_2}{2}\right)^2 + \left(\frac{R_2}{2}\right)^2 \right) + \frac{Q_d^2}{4\pi\varepsilon_0} \frac{e^{-\kappa R_2}}{R_2} \tag{2}$$

where $R_2$ is the separation distance between the two particles. The equilibrium separation distance $R_{eq}$ is determined by setting $U' = 0$,

$$U'_2\big|_{R_2=R_{eq}} = \frac{1}{2}m_d\omega_0^2 R_{eq} - \frac{Q_d^2}{4\pi\varepsilon_0}\left(\frac{1}{R_{eq}^2} + \frac{\kappa}{R_{eq}}\right)e^{-\kappa R_{eq}} = 0 \tag{3}$$

while the breathing frequency is found by defining $U''\big|_{R_{eq}} = \frac{1}{2}m_d\omega_{br}^2$,

$$\omega_{br}^2 = \omega_0^2 + \frac{Q_d^2}{4\pi\varepsilon_0}\left(\frac{2}{R_{eq}^3} + \frac{2\kappa}{R_{eq}^2} + \frac{\kappa^2}{R_{eq}}\right)\frac{e^{-\kappa R_{eq}}}{m_d/2} \tag{4}$$

yielding a relationship between $\omega_{br}$ and $\omega_0 = \omega_{sl}$, the sloshing frequency. As shown, both $\omega_{sl}$ and $\omega_{br}$ are functions of the screening parameter $\kappa$ and the dust charge $Q_d$.

3. Experiment and results

The experiment described here was conducted within one of two modified Gaseous Electronics Conference (GEC) rf reference cells located in the CASPER lab at Baylor University [6]. An open-ended glass box having dimensions of 12 mm × 10.5 mm (height × width) and a glass wall thickness of 2 mm was placed on the powered lower electrode to provide enhanced horizontal confinement for the dust particles (Fig. 1).

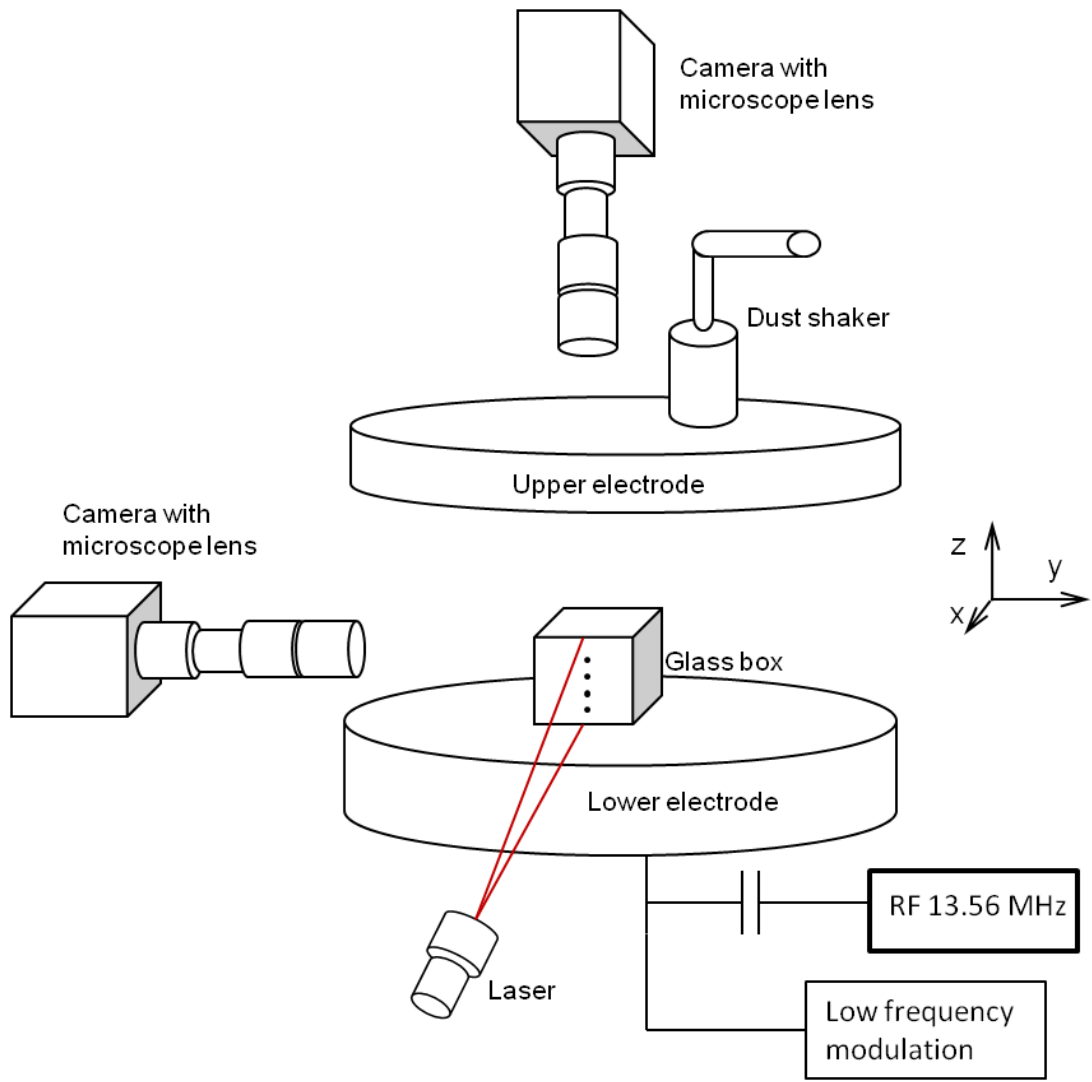

Fig 1. Experimental setup. The open-ended glass box shown here has dimensions of 12 mm × 10.5 mm (height × width). Dust particle oscillation was generated through modulation of an external DC bias introduced on the lower electrode.

Plasma was produced in Argon gas having a neutral gas pressure of 20 Pa by providing 4 W of rf power. Melamine formaldehyde dust particles 8.89 ± 0.09 μm in diameter were introduced into the plasma via shakers mounted above the hollow upper electrode. Under these conditions, dust particles confined inside the glass box formed a turbulent cloud. Slowly lowering the rf power caused the dust cloud to 'stretch' in the vertical direction while simultaneously 'shrinking' in the horizontal direction, reducing the total number of trapped particles through particle loss to the lower electrode. Lowering the rf power further then created a single vertical dust chain. Once stable, the length of this dust chain could be shortened by slightly lowering the rf power, which caused the lowest particle in the chain to enter a region with an unstable force balance, removing it from the chain [6]. On the other hand, by slowly increasing the rf power the overall structural configuration of the chain can be changed starting from a single vertical chain, through a two-dimensional zig-zag structure, to multiple chains exhibiting three-fold symmetry or above. One of the more interesting structures exhibiting three-fold symmetry is the helical structure shown below. See Fig 2(a – e).

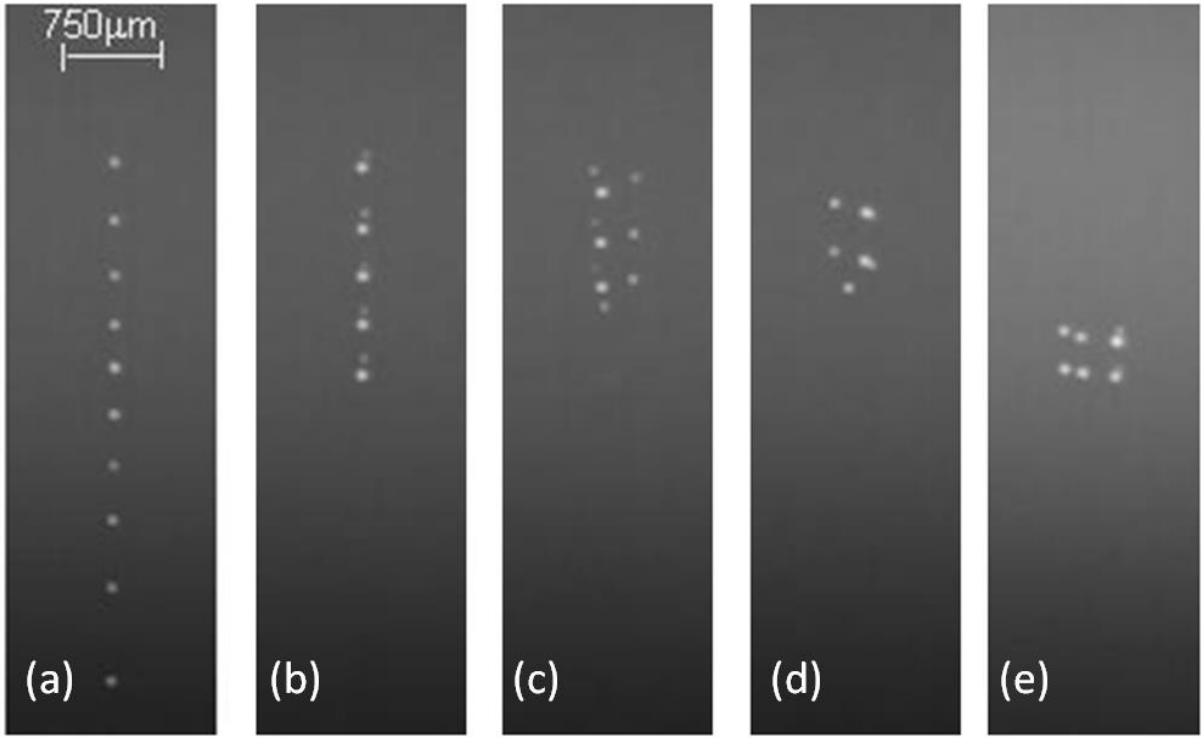

Fig 2. Side view of helical structures formed from 10 particles at varying rf powers using the technique described in the text. In all cases, the background pressure is held at 20 Pa. (a) 1D single chain, rf power 1.23 W. (b) 2D zigzag structure, rf power 1.27 W. (c) 3D three-chain helical structure, rf power 1.32 W. (d) 3D four-chain helical structure, rf power 1.39 W. (e) 3D five-chain helical structure, rf power 1.80 W. Due to the width of the laser sheet used for illumination and the specific portion of the chain imaged by the camera, not every particle in each structure is visible in these side view images.

As stated above, a two-particle vertical chain was chosen to test the validity of Eqs 3 and 4, due to its stable structure during forced oscillation under different rf powers. Fig 3 shows the relative vertical positions and interparticle separations for such a two-particle chain under various rf powers.

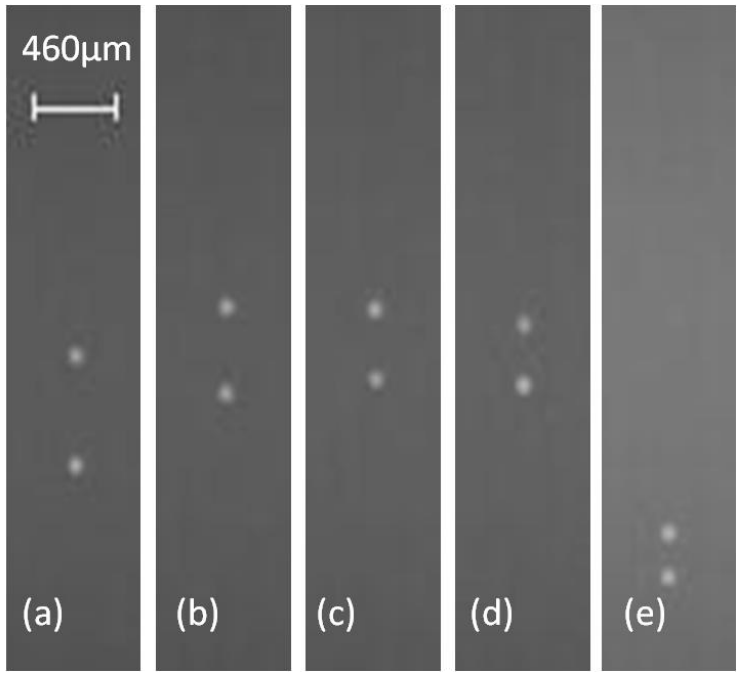

Fig 3. Vertical height and interparticle separation variations for a two-particle chain under various rf powers. From (a) to (e) the corresponding rf powers are 1.23, 1.27, 1.32, 1.39 and 1.80 W, respectively. These power settings are identical to those used for transitioning from a single-chain through a five-chain structure when ten particles are confined within the glass box (Fig 2).

Forced oscillation of the two-particle chain was achieved by modulating the DC bias on the lower electrode employing an externally generated, low frequency sinusoidal signal. The driving amplitude of this signal was limited to 100 mV (peak to peak) measured before a 50 dB attenuator in order to ensure all particle oscillations remained in the linear regime. A side mounted camera collected data on vertical chain oscillations at 125 frames per second (fps). Fig 4 shows the frequency response spectra for both the sloshing and breathing modes occurring at 1.23 W rf power from driven oscillations. At all power settings, the peak sloshing amplitude is greater than that observed for the breathing mode. The relative difference between the two modes increases as rf power increases (Fig 5a). This is in good agreement with mode analysis from data produced by the dust thermal motion which shows the same results (Fig 5b) [7 – 9].

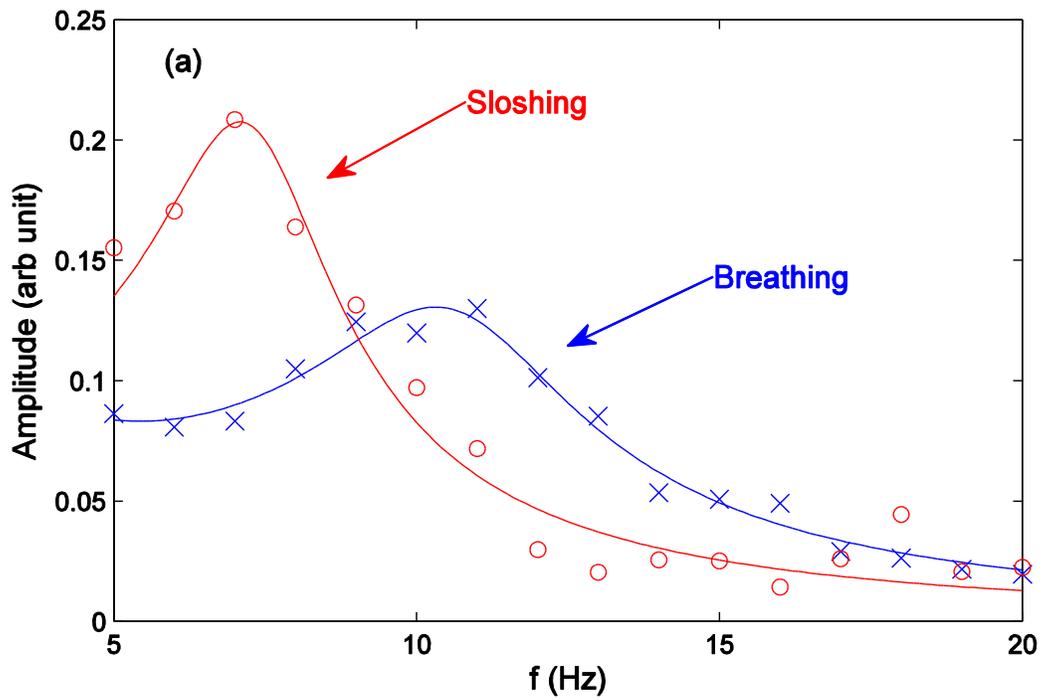

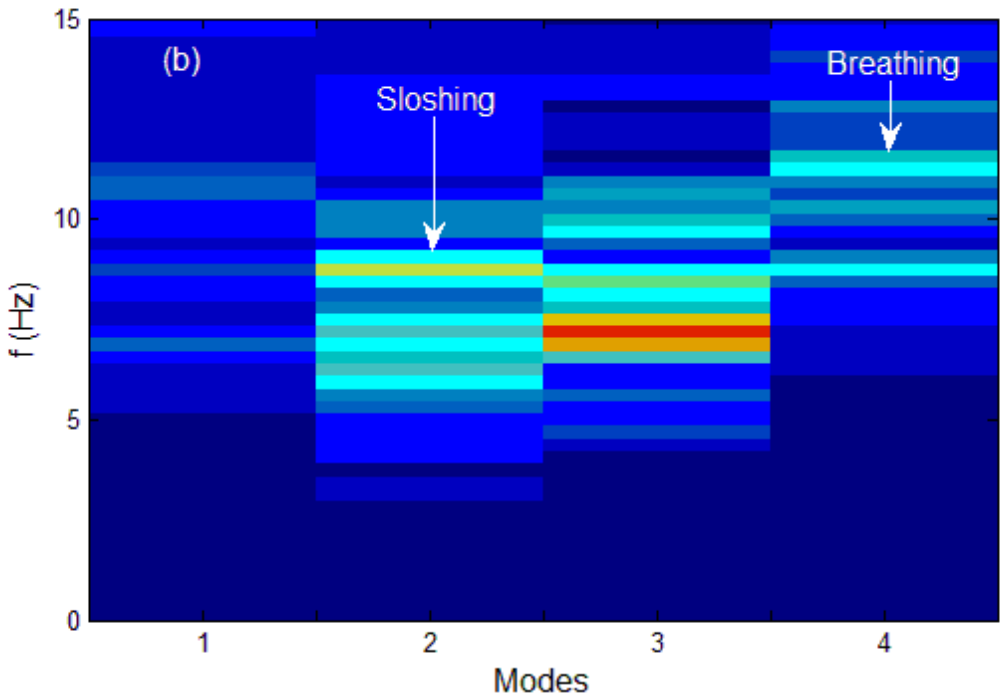

Fig 4. (a) Frequency response data for vertically driven oscillation data for a two-particle chain at a rf power of 1.23 W. Symbols represent experimental data while solid lines represent the theoretical fit as explained in the text. (b) Mode analysis showing the same values for sloshing and breathing oscillation frequencies as derived from the driven

oscillation method. The four modes shown are the 1) horizontal relative, 2) vertical sloshing, 3) horizontal sloshing and 4) vertical breathing modes.

The modulation of the driving frequency was varied from 1 Hz to 30 Hz in order to obtain the frequency response spectra necessary for analysis. The theoretical fit to the experimental data shown in Fig 4a is determined employing the formula for the amplitude of a damped, driven oscillator [10, 11].

$$A = \frac{a_0}{\sqrt{\left(\omega_0^2 - \omega^2\right)^2 + \beta^2 \omega^2}} \quad (5)$$

where $a_0$ is a constant, $\beta$ is the neutral drag coefficient, $\omega_0$ is the resonance frequency (i.e., depending on the reference system $\omega_0$ can represent either the sloshing or breathing frequency), and $\omega$ is the driving frequency. $a_0$, $\omega_0$ and $\beta$ are all determined by fit to the experimental data.

Experimental results for the breathing and sloshing frequencies, their ratio and the equilibrium separation as a function of the rf power can each be determined experimentally and are shown in Fig 5.

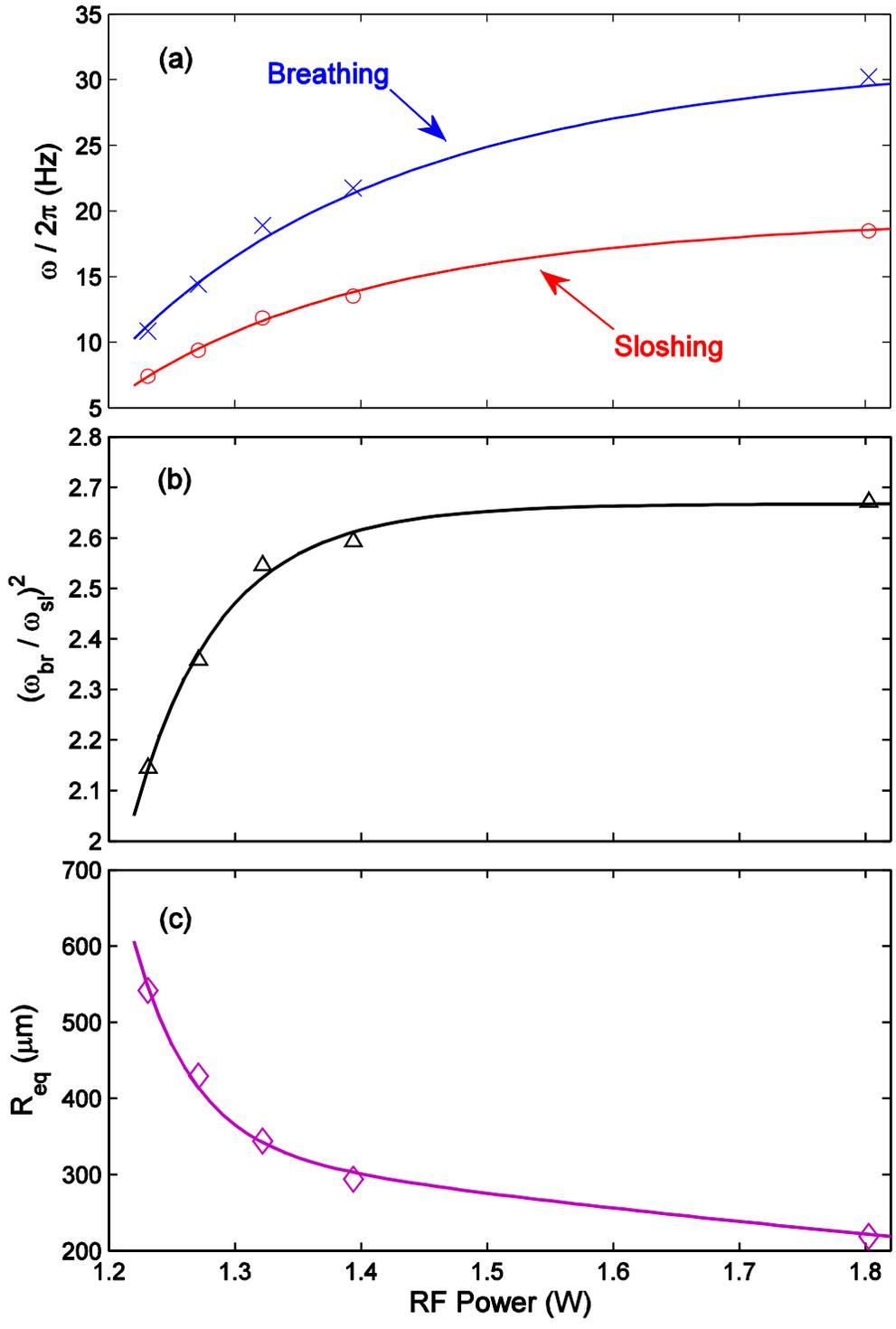

Fig 5. (a) Breathing and sloshing frequencies, (b) the square of the ratio of the breathing to sloshing frequency as a function of the rf power, and (c) the equilibrium separation distance between the two particles as a function of the rf power. Solid lines are included to guide the eye.

As shown, both breathing and sloshing frequencies increase as the rf power increases. While the square of the ratio of the two frequencies squared also increases with rf power, the range is bounded by $2 < (\omega_{br}/\omega_{sl})^2 < 2.7$. Both the frequency ratio and $R_{eq}$ approach limiting values as the power is increased.

4. Analysis

Eqs 3 and 4 are functions both of experimentally determined values, $R_{eq}$, $\omega_{sl}$ and $\omega_{br}$, and unknown values, $Q_d$ and $\kappa$. For $\kappa \to 0$ (the Coulomb interaction limit), a relationship between the breathing and sloshing frequency can be derived,

$$\omega_{br}^2 = 3\omega_{sl}^2 \tag{6}$$

setting a lower limit on the breathing frequency. Fig 5(b) indicates that all experimentally determined values for the frequency ratio $(\omega_{br}/\omega_{sl})^2 < 2.7$ across the rf power settings used, indicating that Eqs 3 and 4 are incomplete.

Assuming small values of $R_{eq}$, this discrepancy between experimental results and theoretical formulation may be corrected by adding a linear force term, $-K_0 R_{eq}$, to Eq 3, where $K_0$ has yet to be identified. Eqs 3 and 4 may now be re-written as,

$$U_2'\big|_{R_2=R_{eq}} = \frac{1}{2}m_d\omega_0^2 R_{eq} - \frac{Q_d^2}{4\pi\varepsilon_0}\left(\frac{1}{R_{eq}^2} + \frac{\kappa}{R_{eq}}\right)e^{-\kappa R_{eq}} - K_0 R_{eq} = 0 \tag{7}$$

$$\omega_{br}^2 = 2U_2''/m_d = \omega_0^2 + \frac{Q_d^2}{4\pi\varepsilon_0}\left(\frac{2}{R_{eq}^3} + \frac{2\kappa}{R_{eq}^2} + \frac{\kappa^2}{R_{eq}}\right)\frac{e^{-\kappa R_{eq}}}{m_d/2} - 2K_0/m_d \tag{8}$$

Unfortunately, with this addition there are now three unknowns, $K_0$, $Q_d$ and $\kappa$ in Eqs 7 and 8. This issue can be ameliorated in part by examining the system in the horizontal plane.

In the horizontal direction, dust particles arrive at equilibrium upon achieving a balance between the confinement produced by the glass walls and the Yukawa interaction between neighboring particles. Therefore, assuming the glass walls exert a horizontal Yukawa confining force to the dust particles, at equilibrium this confining force will be balanced by the horizontal component of the interaction force produced by the dust particles themselves,

$$F_0 \sum_k \frac{\exp(-\kappa R_k)}{R_k^2} = \frac{Q_0^2}{4\pi\varepsilon_0}\sum_{j\neq 0}\frac{\exp(-\kappa r_{0j})}{r_{0j}^2} \tag{9}$$

In Eq 9, the left-hand-side represents the horizontal force produced by the glass walls, with $R_k$ defined as the distance between the $k^{th}$ wall and the representative dust particle, $r_0$. The right-hand-side represents the horizontal component of the interaction force between the dust particle and all neighboring particles. For clusters with small total numbers of particles (for example $N \leq 12$ as shown in the following figure) at the same rf power $F_0$ and $Q_0$ can be assumed constant, i.e., $F_0$ and $Q_0$ are functions of the rf power alone. Representative clusters and corresponding rf powers are shown in Fig 6.

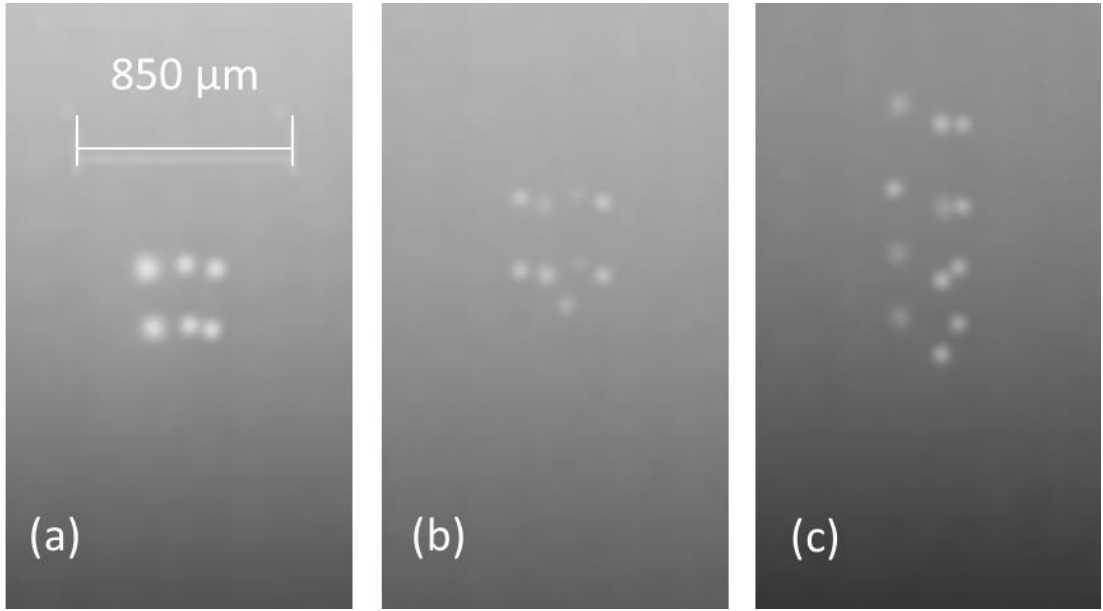

Fig 6. Small clusters with (a) six particles at 1.69 W, (b) nine particles at 1.47 W and (c) twelve particles at 1.36 W.

In this case, particle positions are measured experimentally for clusters having six, nine and twelve particles and formed at 1.36 W, 1.47 W and 1.69 W rf power. Applying Eq 9 to these clusters at the same rf power, the screening parameters may be determined as shown in Fig 7.

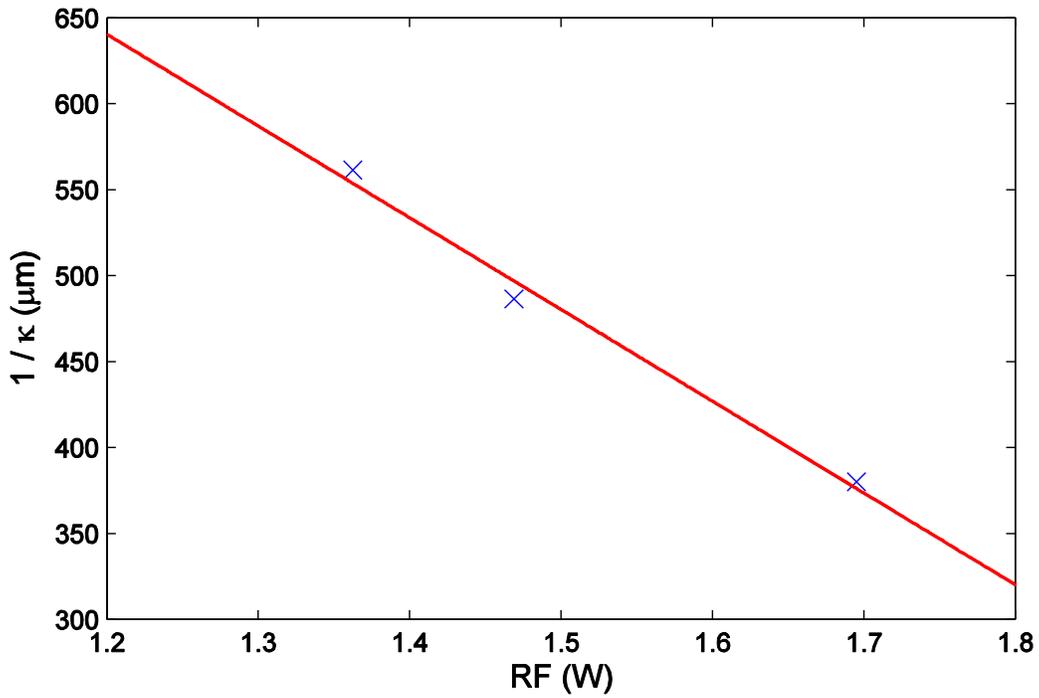

Fig 7. Experimental result of the screening parameter as a function of the rf power.

Using a linear fit for the Debye length as a function of the rf power, the Debye length for the conditions shown in Fig 2a – e are given in Table 1. Substituting these values into Eqs 7 and 8 allows the dust charge $Q_d$ and linear force constant $K_0$ to be derived. These are shown in Fig 8.

Table 1

| RF Power (W) | 1.23 | 1.27 | 1.32 | 1.39 | 1.80 |
|---|---|---|---|---|---|
| $\lambda_D$ (µm) | 624 | 603 | 576 | 539 | 320 |

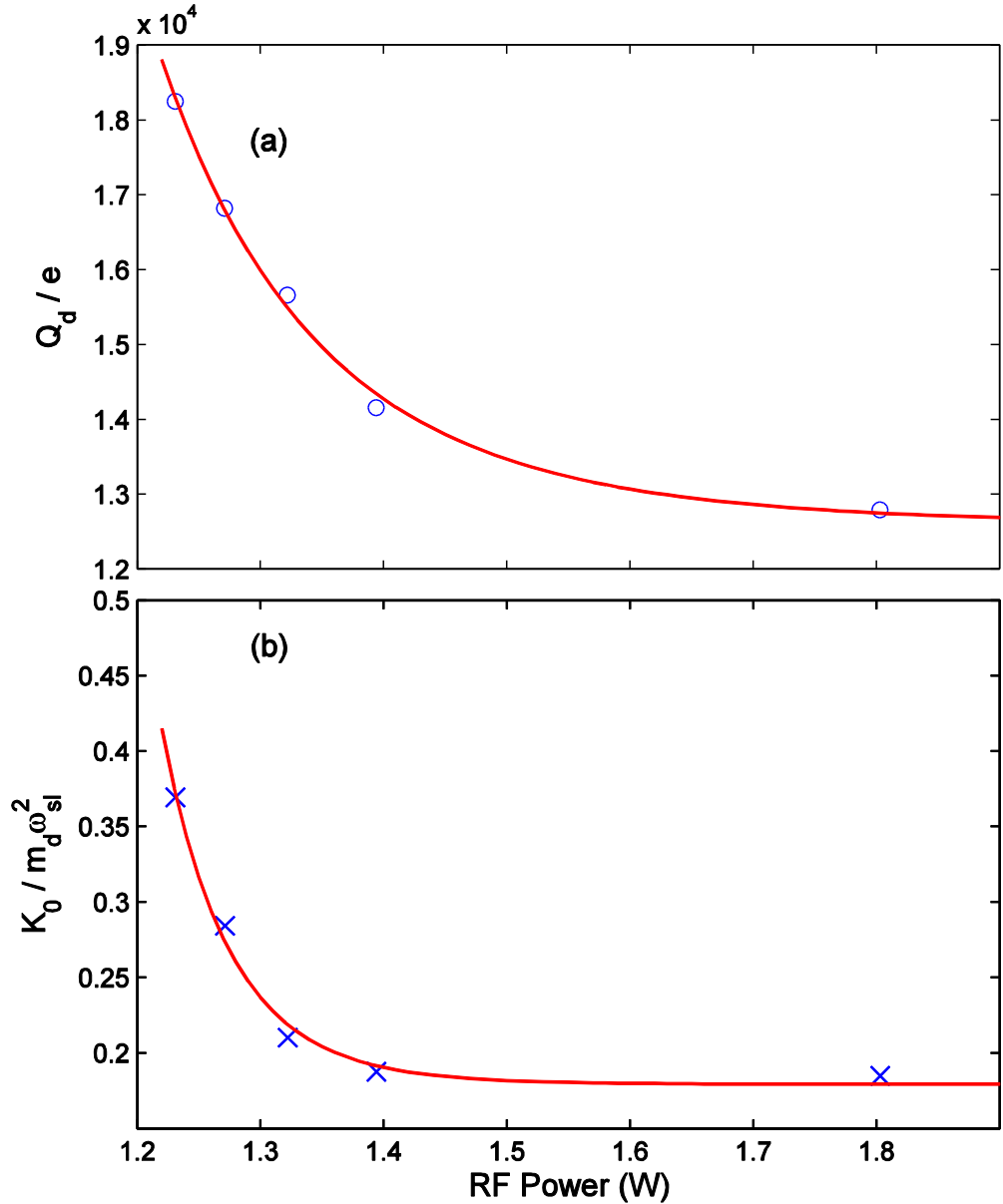

Fig 8. (a) Dust charge, and (b) ratio of the linear force constant $K_0$ to $m_d \omega_{sl}^2$ as a function of the rf power. Fit lines serve to guide the eye.

5. Discussion

As can be seen in Fig 7, the Debye length, $\lambda_D = 1/\kappa$, decreases as the rf power increases. Given the Debye length is inversely related to the plasma density, such a decrease in Debye length and increase in electron and ion densities can best be explained by the increase in ionization occurring with increasing rf power.

Therefore, a *decrease* in rf power will produce an *increase* in Debye length, providing less shielding to the dust particles within the box from the potential on the glass walls comprising the box. (This assumes a Yukawa type confinement force.) This is the primary reason for the structural transition from multi-chain to single chain, which is observed as the rf power decreases as shown in Fig 2.

Fig 8b shows that the linear force term constant $K_0$ as defined for Eqs 7 and 8 remains relatively uniform at approximately 20% of the value of the sloshing frequency, $m_d \omega_{sl}^2$, over the rf power regime where multiple chains form. However, as the rf power decreases below 1.4 W, the rf power regime where single chains form, $K_0$ rapidly increases to 40% of the sloshing frequency. As such, this force provides a significant contribution to Eqs 7 and 8.

One possible reason for this result has to do with the potential on the surface of the glass walls. The force produced by the glass walls establishes much of the horizontal confinement acting on the dust particles. However, previous experiments have shown that the glass box also provides a vertical force on the dust particles [11]. This vertical force from the glass box is the primary contributor to the linear term $-K_0 R_{eq}$.

A second possible contributor is the ion focusing effect. The wake effect due to streaming ions has been shown to create a charge difference between two vertically aligned dust particles and to create a positive charge space region below the upper particle [12]. Thus the ion focusing effect introduces an additional attractive force between the dust particles [13, 14].

Finally, it has been noted that ion-neutral charge exchange collisions can lead to trapped ions around dust grains, which reduces the grain charge substantially in laboratory plasmas due to the increase in ion-neutral collisions with increasing plasma density (15, 16, 17, 18). However, all previous studies examined only the effect due to trapped ions in an isotropic plasma, neglecting the effects of flowing ions or gradients in the electric potential, both of which exist in the plasma sheath. The fact that the measured charge on a dust grain in this experiment decreases with increasing plasma power indicates that such collisional effects may well play a significant role in grain charging, even in the sheath.

6. Conclusions

The total potential energy of a vertically aligned, two-particle dust cluster system and its first and second derivatives were employed to investigate the relationship between the system's sloshing and breathing frequencies. These were then employed to determine the Debye length and dust charge as a function of the rf power.

In order to model the relationship between the frequencies of the breathing and sloshing modes, an additional linear force term (see Eqs 7 and 8), $-K_0 R$, was introduced to solve the discrepancy between the values predicted theoretically, (see Eqs 3 and 4) and the experimental results. The origin of this linear force term was examined and two possible mechanisms, the glass box potential and ion focusing effect, were proposed.

The experimental results provided show that the Debye length decreases as the rf power increases, explaining the transition observed for dust particle bundles confined within a glass box from vertically aligned single chains to multiple chains as the rf power increases.

The relationship between the dust charge and the system rf power was also examined and is shown in Fig 8a. These results indicate that ion-neutral collisions can lead to trapped ions around the dust grains and play a significant role in the grain charging process inside the glass box.